\documentclass[10pt]{article}

\usepackage{amsfonts} 
\usepackage{amsmath,hhline,epsfig,latexsym}
\usepackage{amssymb}
\usepackage{hyperref}
\usepackage{enumerate}

\newtheorem{theorem}{\sc Theorem}[section]

\newcommand{\N}{\mbox{$I \kern -4pt N$}}
\newcommand{\Q}{\mbox{$Q \kern -8pt I$}}
\newcommand{\R}{\mbox{$I \kern -4pt R$}}
\newcommand{\C}{\mbox{$C \kern -8pt I$}}

\def\R{{\bf R}}

\begin{document}

\title{\textbf{Robust Optimal Design of Quantum Electronic Devices}}

\author{\textsc{Ociel Morales,}\thanks{%
Facultad de Ciencias. Universidad Aut\'onoma de San Lu\'{\i}s Potos\'{\i} (M\'exico). Email: \texttt{omorales@fc.uaslp.mx}}\quad \textsc{Francisco Periago}\thanks{%
Departamento de Matem\'{a}tica Aplicada y Estad\'{i}stica. Universidad Polit\'ecnica de Cartagena (Spain). Email: \texttt{f.periago@upct.es}}  \quad and\quad \textsc{Jos\'e A. Vallejo}\thanks{%
 Facultad de Ciencias. Universidad Aut\'onoma de San Lu\'{\i}s Potos\'{\i} (M\'exico). Email: \texttt{jvallejo@fc.uaslp.mx}}}

\maketitle

\begin{abstract}

We consider the optimal design of a sequence of quantum barriers in order to manufacture an
electronic device at the nanoscale such that the dependence of its transmission coefficient 
on the bias voltage is linear. The technique presented here is easily adaptable to other 
response characteristics.
The transmission coefficient is computed using the Wentzel-Kramers-Brillouin (WKB) method, so
we can explicitly compute the gradient of the objective function. In contrast with
earlier treatments, manufacturing uncertainties are incorporated in the model through random
variables and the optimal design problem is formulated in a probabilistic setting. As a 
measure of robustness, a weighted sum of the expectation and the variance of a least-squares
performance metric is considered. Several simulations illustrate the proposed approach.
\end{abstract}

\noindent
\textbf{Keywords:} Robust Optimal Design, Nanoelectronics,  Stochastic Collocation Methods, WKB Approximation.

\section{Introduction}

Nanoelectronic devices operate with extremely low intensity currents. Under these circumstances, it 
is desirable to have at our disposal mechanisms to produce and control electronic currents with a
high precision. Electronic beams are relatively easy to produce, but their filtering to obtain
nanocurrents with specified properties is much more difficult. A widely used approach consists in
directing the beam on a sequence of quantum barriers with an externally adjustable bias voltage 
applied throughout the device. One expects to be able to control the response of the device in the
form of a current whose intensity depends, say, linearly on the applied bias. This setting naturally
leads to an optimal design problem: What must be the width and height of the layers composing the
barriers, supposed fixed in number, in order to achieve this linear response? (Of course, the problem 
is quite general, admitting a more complex relation between the external voltage and the current, but 
here we deal with the linear  case just for simplicity).\\
There should be no need to stress the importance of the solution to this problem from a practical 
point of view, but it must be noticed right from the start that a closed-form, analytic solution is
impossible to obtain in most cases. 
The use of numerical computations at some stage is unavoidable, and this 
leads to the question of which method to use in order to obtain a good approximation to the solution.
In \cite{Levi2010} the non--constant potential energy profile is approximated by piecewise constant potentials. Then, the propagation matrix method \cite{Gilmore_book,Levi_book} is applied to compute the transmission coefficient and, finally, the gradient of a least--squares--type  objective function (which is required by the numerical solution method)  is computed using the adjoint method.  It is important to point out that different discretization processes, which are used to approximate objective functions and/or its gradients, may lead to very different results. Moreover,
it has been observed in some optimization problems \cite{hager} that first approximating a cost functional, and then computing the gradient of the approximated one, in general differs from approximating the gradient of the exact cost functional. That is, the schemes `first discretize, then optimize' and `first optimize, then discretize', do not commute in general.  Also, as it will be showed later on in this paper, optimizing for the same cost functional via its exact gradient gives different solutions than using an approximate one.

Another issue, which cannot be obviated in a realistic mathematical model, is the presence of uncertainties. There are several sources of uncertainty in the problem under consideration,
one of the most important regarding the influence on the computed optimal design being the
manufacturing uncertainties. Due to the smallness of the currents involved, and the narrow width
of the quantum barriers needed, methods such as MBE (Molecular Beam Epitaxy) or CVD (Chemical
Vapor Deposition) are used to growth thin layers (in many cases, monolayers) of some material
to build the barriers, two of the preferred ones being $MoS_2$ and $GaAs$ (see \cite{bergeron,Gong}, 
for example). These methods allow the growth even of monolayers, but the difficulties inherent
to the manufacturing process at a semi--commercial scale lead almost inevitably to inaccuracies
that ultimately lead to a potential configuration that may be different from the numerically
computed, optimal one \cite{Schmidt2006}.   
For these reasons, the problem of computing an optimal quantum profile which, in addition, is robust against those uncertainties is an important one. If there is some statistical information about the uncertainties, then the machinery of probability theory gives a framework in which to we can accomodate uncertainties (by using random variables and/or random fields), and model objective functions (by means of expectation and variance operators, among others choices). In \cite{Zhang2007}, this approach has been used for the case in which the cost functional only includes the averaging
of a least--squares performance metric, and by using the standard Monte--Carlo method for its numerical resolution.

The present work addresses the problem of the optimal design of a quantum potential profile
(modeling a nanoelectronic device) in order to obtain a transmission coefficient linearly depending
on an externally applied bias voltage, in the presence of manufacturing uncertainties.
The transmission coefficient is explicitly computed by using the WKB method. 
As a consequence, an explicit  formula for the gradient of the cost functional is obtained. 
A weighted sum of expectation and  variance of a random least--squares performance metric is considered as a measure of robustness. The inclusion of the second order statistical moment in the cost functional amounts to a reduction the dispersion of the random transmission coefficient and hence an increase in the robustness of the optimal design.  Since the resulting integrand in the cost functional is smooth with respect to a random parameter, a sparse grid stochastic collocation method is used for the numerical approximation of the involved integrals in the random domain. This method preserves the parallelizable character of Monte--Carlo sampling. However, in contrast to Monte--Carlo (which is computationally very expensive, of order $O\left( M^{-1/2}\right)$, with $M$ the number of random sampling points), the stochastic collocation method shows an exponential convergence with respect to the number of sampling points. Several simulations illustrate the proposed approach, which shows itself to be an improvement in accuracy over previous ones of about a $69{.}4\%$.

\section{Setting of the Optimal Design Problems}

Considered is a nanoscale semiconductor electronic device composed of $N$ layers occupying positions $x_0=0<x_1<\cdots<x_N=L$. The local potential energy at the $i$th layer is denoted by $U_i$, $i=1,2,\cdots,N$. For $x<x_0$, the potential energy is denoted by $U_0$ and for $x>x_N$ it is $U_{N+1}$. It is assumed that a single electron propagating from $-\infty$ is incident at $x_0$ and that a voltage bias $V_{\text{bias}}$ is applied across the device. A linear approximation of the underlying Poisson's equation \cite{Levi2010,Wang_book} leads to the following expression for the resulting potential energy profile
\begin{equation}\label{potential_V}
V\left( x\right)=V\left( x,U,V_{\text{bias}}\right)=\left\{
\begin{array}{ll} 
 U_0,  &  -\infty<x<x_0\\
\displaystyle\sum_{j=1}^NU_j\mathcal{X}_j\left( x\right)-V_{\text{bias}}\frac{x-x_0}{L} , & x_0\leq x\leq x_N \\
U_{N+1}-V_{\text{bias}}, & x_N<x<+\infty ,
\end{array}
\right.
\end{equation}  
where $U=\left( U_1,\cdots,U_N\right)$ is the vector of local layer potentials in the device, and $$\mathcal{X}_j\left( x\right) = \left\{
\begin{array}{ll} 
 1,  &  x_{j-1}\leq x<x_j\\
0, & \text{otherwise} ,
\end{array}
\right.$$
is the characteristic function of the interval $[x_{j-1},x_j[$, $1\leq j\leq N$.

The transmission coefficient of the device $T=T\left( V_{\text{bias}},U\right)$ is defined as the ratio of current density transmitted from the device at $x=x_N$ and the incident one at $x=x_0$. As explained in detail in \cite{Levi2010}, $T$ may be expressed as

\begin{equation}
\label{coef_transmission}
T\left( V_{\text{bias}},U\right) = \frac{\kappa_{N+1}}{\kappa_{0}}\vert \psi\left( x_N\right)\vert\,,
\end{equation}
where $\kappa_{0}=\sqrt{2me\left(E-U_0\right)}/\hslash$ and $\kappa_{N+1}=\sqrt{2me\left(E-U_{N+1}+V_{\text{bias}}\right)}/ \hslash$, for values of the energy $E>\max\left\{ U_0, U_{N+1}-V_{\text{bias}}\right\}$. The cases $E\leq U_0$ and $E\leq U_{N+1}-V_{\text{bias}}$ may be treated in a similar way.  Here  $m$ is the effective mass of the electron, $e$ denotes the electron charge, $\hslash$ is Planck's constant,  $E$ is the electron energy, and finally $\psi\left( x\right)$ solves the following boundary-value problem for the Schr\"odinger equation
\begin{equation}
\label{schrodinger}
\left\{\begin{array}{ll}
-\dfrac{\hslash^2}{2m}\dfrac{d^2\psi\left( x\right)}{dx^2}+V\left( x\right)\psi\left( x\right)=E\psi\left( x\right), & x_0<x<x_N \\[7pt]
\mathbf{i}\kappa_{0}\psi\left( x_0\right)+\dfrac{d\psi}{dx}\left( x_0\right) = 2\mathbf{i}\kappa_{0}A_0e^{\mathbf{i}\kappa_{0}x_0}, \\[7pt]
\mathbf{i}\kappa_{N+1}\psi\left( x_N\right)-\dfrac{d\psi}{dx}\left( x_N\right) =0\,.
\end{array}
\right.
\end{equation}
Here $\mathbf{i}$ denotes the unit imaginary number and $A_0$ the amplitude of the transmitted wave at $x_0$.  

\subsection{Deterministic Optimal Design}\label{Design_problem}

The (deterministic) optimal design problem considered in this paper is formulated as the following nonlinear data-fitting problem: Given a desired transmission coefficient $T_0\left( V_{\text{bias}}\right)$, which is defined for $V_{\text{min}}\leq V_{\text{bias}}\leq V_{\text{max}}$, and lower, $U_L$, and upper, $U_R$, bounds for the local layer potentials, with $0\leq U_L<U_R<\infty$,
\begin{equation}
\label{ODP_deterministic}
\left\{\begin{array}{ll}
\mbox{Minimize in } & U=\left( U_1,\cdots,U_N\right):  J\left( U\right) = \displaystyle\sum_{i=1}^M\vert T_0\left( V_i\right)-T\left( V_i,U\right)\vert^2 \\
\mbox{subject to} & U_L\leq U_j\leq U_R$,\quad $j=1,\cdots,N\,,
\end{array}
\right.
\end{equation}
where $T\left( V_i,U\right)$ is given by (\ref{coef_transmission}) with $V_{\text{bias}}=V_i$ and $V_{\text{min}}\leq V_i\leq V_{\text{max}}$, $i=1,\cdots,M$.

\subsection{Optimal Design Under Manufacturing Uncertainties}

As indicated in the introduction, it is very convenient to analyse the robustness of optimal designs with respect to  manufacturing uncertainties. These  may be modelled by adding a vector of random variables 
\begin{equation}
\label{uncertain_manufacturing}
X\left( \omega\right)=\left(X_1\left( \omega\right),\cdots,X_N\left( \omega\right)\right)
\end{equation}
to the vector $U$ of local layer potentials. Here $\omega$ represents a random event and thus $X_j\left( \omega\right)$ is a small unknown error in manufacturing the local potential $U_j$. Hence, the cost functional $J$ considered in problem (\ref{ODP_deterministic}) becomes a random variable  given by
\begin{equation}
\label{random_cost}
J\left( U,\omega\right) = \displaystyle\sum_{i=1}^M\vert T_0\left( V_i\right)-\left(T\left( V_i,U+X\left(\omega\right)\right)\right)\vert^2 .
\end{equation} 
In order to obtain a design of the potential energy profile $U$ less sensitive with respect to  fabrication unknown fluctuations, the new cost functional is considered:  
\begin{equation}
\label{robust_cost}
J_{\alpha}\left( U\right)=\mathbb{E}\left( J\left( U,\cdot\right)\right)+\alpha \text{Var}\left( J\left( U,\cdot\right)\right),
\end{equation}
with $\alpha\geq 0$ a weighting parameter. Here $\mathbb{E}$ and Var denote the expectation and variance operators, respectively. Then, the robust optimization problem is formulated  as
\begin{equation}
\label{RDP}
\left\{\begin{array}{ll}
\mbox{Minimize in }& U=\left( U_1,\cdots,U_N\right): J_{\alpha}\left( U\right)  \\
\mbox{subject to} & U_L\leq U_j\leq U_R$,\quad $j=1,\cdots,N.
\end{array}
\right.
\end{equation}
where $J_{\alpha}\left( U\right)$ is given by (\ref{robust_cost}).

\section{Solving the Optimal Design problems}\label{sec:2}

The numerical resolution of the optimal design problems stated in the preceding section requires the computation of the transmission coefficient \eqref{coef_transmission} and, therefore, the resolution of the boundary-value problem \eqref{schrodinger}. This problem may be numerically approximated by standard numerical methods such as finite differences or finite elements. Another approach is proposed in \cite{Levi2010} where, after approximating the potential $V\left( x\right)$, as given by \eqref{potential_V}, by piecewise constant potentials, problem \eqref{schrodinger} is transformed into a two--dimensional linear non--autonomous difference equation. Here we propose a different approach based on the so--called WKB method \cite{Maslov}. From the point of view of optimization, WKB method is very appealing since, within its range of validity, it provides an explicit form for the solution to \eqref{schrodinger}. From this, explicit expressions for the gradients of the cost functionals considered in problems \eqref{ODP_deterministic} and \eqref{RDP} are derived. In addition, having an explicit expression for  $J\left( U,\omega\right)$  allows us to prove its smoothness with respect to $U$ and $\omega$. From this, both existence of solutions to \eqref{ODP_deterministic} and \eqref{RDP}, as well as designing a computationally very efficient numerical resolution method, will be derived in this section.

We begin by explicitly computing the transmission coefficient \eqref{coef_transmission} and then describe the numerical resolution methods for problems \eqref{ODP_deterministic} and \eqref{RDP}.

\subsection{Explicit Computation of Transmission Coefficient}

\subsubsection{Case of a single potential barrier}

\begin{figure}[h!]
\includegraphics[scale=1]{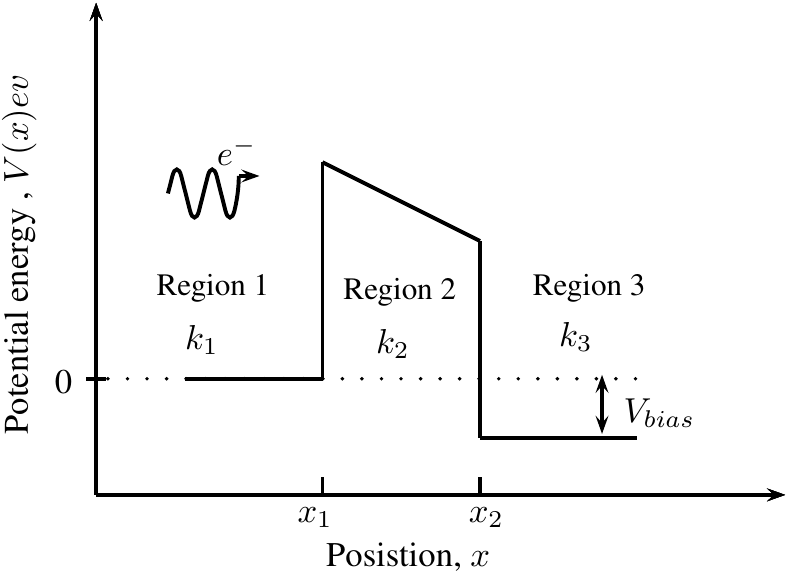}
\centering
\caption{Sketch of the onde-dimensional potential energy barrier $V(x)=U-V_{\text{bias}}\frac{x-x_1}{x_2-x_1}{,} $  $x_1\leq x\leq x_2 {.}$ An electron of mass $m$,  charge $e$ and energy $E $, incident from left, has wave vector $k_j$ in region $j$.}
\label{fig:potential}
\end{figure}

For the sake of clarity, consider first the case of a single potential barrier as illustrated in Figure \ref{fig:potential}. 

The WKB method proposes a solution of the Schr\"odinger equation in the form 
\begin{equation}
\psi(x) =
\begin{cases}
\psi_1(x)=A_1e^{\frac{\mathbf{i}}{\hbar}\kappa_1 x}+B_1e^{-\frac{\mathbf{i}}{\hbar}\kappa_1x},& x<x_1\\
\psi_2(x)=\frac{A_2}{\sqrt{\kappa_2(x)}} e^{\frac{\mathbf{i}}{\hbar}\int_{x_1}^{x}\kappa_2(s)ds}+\frac{B_2}{\sqrt{\kappa_2(x)}} e^{-\frac{\mathbf{i}}{\hbar}\int_{x_1}^{x}\kappa_2(s)ds}, & x_1\leq x\leq x_2\\
\psi_3(x)=A_3e^{\frac{\mathbf{i}}{\hbar}\kappa_3 x}+B_3e^{-\frac{\mathbf{i}}{\hbar}\kappa_3 x},& x>x_2,
\end{cases}
\end{equation}
with $\kappa_1 = \sqrt{2meE}$, $\kappa_2(x)=\sqrt{2me(E-V(x))}$ if $E>V\left( x \right)$ for all
$x_1\leq x\leq x_2 $, and $\kappa_3(x)=\sqrt{2me(E+V_{\text{bias}})}$ for all $x>x_2$.

In the context of quantum electronic devices, solutions to \eqref{schrodinger} admit a smooth
representative in their $L^2$ classes (of regularity class $C^1$). Hence, continuity of $\psi$
and its first derivative at the interface point $x_1$  leads to 
\begin{align}\label{transmission_x_1}
&\begin{bmatrix}
e^{\frac{\mathbf{i}}{\hbar}\kappa_1 x_1} & e^{-\frac{\mathbf{i}}{\hbar}\kappa_1x_1}\\
\frac{\mathbf{i}}{\hbar}\kappa_1e^{\frac{\mathbf{i}}{\hbar}\kappa_1x_1} & -\frac{\mathbf{i}}{\hbar}\kappa_1e^{-\frac{\mathbf{i}}{\hbar}\kappa_1x_1} 
\end{bmatrix}
\begin{bmatrix}
A_1\\
B_1
\end{bmatrix}
 \nonumber\\
&= \begin{bmatrix}
\frac{1}{\sqrt{\kappa_2(x_1)}} & \frac{1}{\sqrt{\kappa_2(x_1)}}\\
 & \\
-\frac{C}{2\kappa_2^2(x_1)\sqrt{\kappa_2(x_1)}} 
+\frac{\mathbf{i} \kappa_2(x_1)}{\sqrt{\kappa_2(x_1)}}
 & -\frac{C}{2\kappa_2^2(x_1)\sqrt{\kappa_2(x_1)}} 
-\frac{\mathbf{i}\kappa_2(x_1)}{\sqrt{\kappa_2(x_1)}}
\end{bmatrix}
\begin{bmatrix}
A_2 \\ B_2
\end{bmatrix}
\end{align}
where 
\begin{equation}\label{constant_C}
C=C\left( V_{\text{bias}}\right)=\frac{meV_{\text{bias}}}{x_2-x_1} .
\end{equation}
By writing each $2\times 2$ matrix in (\ref{transmission_x_1}) as the product of two matrices as follows
\begin{equation}
\label{decomposition_x_1}
\begin{array}{ll}
& \underbrace{
\begin{bmatrix}
1 & 1\\
\frac{\mathbf{i}}{\hbar}\kappa_1 & -\frac{\mathbf{i}}{\hbar}\kappa_1
\end{bmatrix}}_{K\left(\kappa_1\right)}
\underbrace{
\begin{bmatrix}
e^{\frac{\mathbf{i}}{\hbar}\kappa_1x_1} & 0\\
0 & e^{-\frac{\mathbf{i}}{\hbar}\kappa_1x_1}
\end{bmatrix}}_{E\left( \kappa_1, x_1\right)}
\begin{bmatrix}
A_1\\
B_1
\end{bmatrix} \\
& =
\underbrace{ \begin{bmatrix}
1 & 1\\
\frac{-C}{2\kappa_2(x_1)}+\frac{\mathbf{i}\kappa_2(x_1)}{\hbar} & \frac{-C}{2\kappa_2(x_1)}-\frac{\mathbf{i}\kappa_2(x_1)}{\hbar}
\end{bmatrix} }_{\mathcal{K}\left( \kappa_2, x_1\right)}
\underbrace{
 \begin{bmatrix}
 \frac{1}{\sqrt{\kappa_2(x_1)}} & 0 \\
 0 & \frac{1}{\sqrt{\kappa_2(x_1)}}
\end{bmatrix}}_{\mathcal{E}\left( \kappa_2, x_1\right)}
\begin{bmatrix}
A_2 \\ B_2
\end{bmatrix}
\end{array}
\end{equation}
and solving (\ref{decomposition_x_1}) for $A_1$ and $B_1$,
\begin{equation}\label{step1}
\begin{bmatrix}
A_1 \\ B_1
\end{bmatrix}
= E^{-1}(\kappa_1,x_1)K^{-1}(\kappa_1)\mathcal{K}(\kappa_2,x_1)\mathcal{E}(\kappa_2,x_1)
\begin{bmatrix}
A_2 \\ B_2
\end{bmatrix}.
\end{equation}
Proceeding in the same way at the point $x_2$, one obtains
\begin{equation}\label{step2}
\begin{bmatrix}
A_2 \\ B_2
\end{bmatrix}
= \mathcal{E}^{-1}(\kappa_2,x_2)\mathcal{K}^{-1}(\kappa_2,x_2)K(\kappa_3)E(\kappa_3,x_2)
\begin{bmatrix}
A_3 \\ B_3
\end{bmatrix},
\end{equation}
where
\begin{equation*}
\begin{split}
\mathcal{K}(\kappa_j,x)&=
\begin{bmatrix}
1&1\\
-\frac{C}{2\kappa_j^2(x)}+\frac{\mathbf{i}}{\hbar}\kappa_j(x) & -\frac{C}{2\kappa_j^2(x)}-\frac{\mathbf{i}}{\hbar}\kappa_j(x)
\end{bmatrix}
\end{split}
\end{equation*}
and
\begin{equation*}
\begin{split}
\mathcal{E}(\kappa_j,x)&=
\begin{bmatrix}
\frac{e^{\frac{\mathbf{i}}{\hbar}\int_{x_{j-1}}^x \kappa_j(s)\,ds}}{\sqrt{\kappa_j(x)}} & 0\\
0&\frac{e^{-\frac{\mathbf{i}}{\hbar} \int_{x_{j-1}}^x \kappa_j(s)\,ds}}{\sqrt{\kappa_j(x)}}
\end{bmatrix}{.}
\end{split}
\end{equation*}
Substituting \eqref{step2} into \eqref{step1}, we get
\begin{equation}
\begin{bmatrix}
A_1\\ B_1
\end{bmatrix}
=  E^{-1}(\kappa_1,x_1)K^{-1}(\kappa_1)\mathcal{K}(\kappa_2,x_1)\mathcal{E}(\kappa_2,x_1)\mathcal{E}^{-1}(\kappa_2,x_2)\mathcal{K}^{-1}(\kappa_2,x_2)K(\kappa_3)E(\kappa_3,x_2) \begin{bmatrix}
A_3 \\ B_3
\end{bmatrix}.
\end{equation}
Denoting by $M$ the product of matrices from $E^{-1}(\kappa_1,x_1)$ to $E(\kappa_3,x_2)$
the transmission coefficient (\ref{coef_transmission}) takes the form $
T=\vert \frac{1}{M_{11}}\vert^2$, where $M_{11}$ is the first entry of $M$.  More precisely, the following explicit formula for  that transmission coefficient, in the case $E>V\left( x \right)$ for all $x_1\leq x\leq x_2$, is obtained:
\begin{equation}\label{coef_trans_WKB_E_mayor}
\begin{array}{ll}
T\left(V_{\text{bias}},U \right)&=\frac{\kappa_2(x_1)}{\kappa_2(x_2)}\left[\left(
\frac{\hbar  C \cos \frac{I}{\hbar} }{4 \kappa_1 {{\kappa_2^2(x_1)}} }-\frac{\hbar  \kappa_2(x_1) C \cos \frac{I}{\hbar} }{4 \kappa_1 {{\kappa_2^3(x_2)}} }-\frac{\kappa_2(x_1) \sin \frac{I}{\hbar} }{2 \kappa_1}    
-  \frac{{{\hbar}^{2}} {{C}^{2}} \sin \frac{I}{\hbar} }{8 \kappa_1 {{\kappa_2^2(x_1)}} {{\kappa_2^3(x_2)}} }-\frac{\kappa_3 \sin \frac{I}{\hbar} }{2 \kappa_2(x_2)}    \right)^2 \right. \\
& \left. +{{\left( \frac{\cos \frac{I}{\hbar} }{2}+\frac{\kappa_3 \kappa_2(x_1) \cos \frac{I}{\hbar} }{2 \kappa_1 \kappa_2(x_2)}-\frac{\hbar  C \sin \frac{I}{\hbar} }{4 {{\kappa_2^3(x_2)}}}+\frac{\hbar  C \kappa_3 \sin \frac{I}{\hbar} }{4\kappa_1 {{\kappa_2^2(x_1)}} \kappa_2(x_2)}\right) }^{2}}\right]^{-1},
\end{array}
\end{equation}
where $C$ is given by (\ref{constant_C}) and 
\begin{equation}
\label{I}
I=I\left( V_{\text{bias}},U \right)=  \frac{2\sqrt{2me}(x_2-x_1)}{3mV_{\text{bias}}}\left[ \left( E-U+V_{\text{bias}}\right)^{3/2}-\left(  E-U\right)^{3/2}\right] 
\end{equation}
The  case $E<V\left( x \right)$ for all $x_1\leq x\leq x_2$  is completely analogous. Denoting by 
\begin{equation}\label{kappa_barra}
\overline{\kappa}_2\left( x\right)=\sqrt{2me\left( V\left( x\right)-E\right)},  \quad x_1\leq x\leq x_2,
\end{equation}
the transmission coefficient is given by
\begin{equation}\label{coef_trans_WKB_E_menor}
\begin{array}{ll}
T\left(V_{\text{bias}},U \right)&=\frac{\overline{\kappa}_2(x_1)}{\overline{\kappa}_2(x_2)}
\left[ \left(
\frac{\overline{\kappa}_2(x_1) \sinh \frac{\overline{I}}{\hbar} }{2 \kappa_1}-\frac{\hbar C \cosh \frac{\overline{I}}{\hbar}  }{4 \kappa_1 {{\overline{\kappa}_2^2(x_1)}} }+\frac{\hbar  C\overline{\kappa}_2(x_1) \cosh \frac{\overline{I}}{\hbar} }{4 \kappa_1 {{\overline{\kappa}_2^3(x_2)}} }
-\frac{\hbar^2  C^2\sinh \frac{\overline{I}}{\hbar}}{8 \kappa_1 \overline{\kappa}_2^2(x_1) \overline{\kappa}_2^3(x_2) }
-\frac{\kappa_3\sinh \frac{\overline{I}}{\hbar} }{2 \overline{\kappa}_2(x_2)} \right)^2 \right. \\
& \left. + {\left( \frac{\cosh \frac{\overline{I}}{\hbar} }{2}+\frac{\overline{\kappa}_2(x_1) \kappa_3 \cosh \frac{\overline{I}}{\hbar} }{2 \kappa_1 \overline{\kappa}_2(x_2)}+\frac{\hbar C\sinh \frac{\overline{I}}{\hbar} }{4 {{\overline{\kappa}_2^3(x_2)}}}-\frac{\hbar C \kappa_3 \sinh \frac{\overline{I}}{\hbar} }{4 \kappa_1 {{\overline{\kappa}_2(x_1)}^{2}} \overline{\kappa}_2(x_2)}\right) }^{2}
\right]^{-1}
\end{array}
\end{equation}
where 
\begin{equation}\label{I_barra}
\overline{I} = \overline{I}\left( V_{\text{bias}},U\right) = \frac{2\sqrt{2me}(x_2-x_1)}{3mV_{\text{bias}}}\left[ \left( U-E\right)^{3/2}-\left( U-E-V_{\text{bias}}\right)^{3/2}\right] 
\end{equation}

\subsubsection{Range of validity for WKB method }

\begin{figure}[h!]
\centering
	    \begin{tabular}{c c}
	    \includegraphics[scale=0.51]{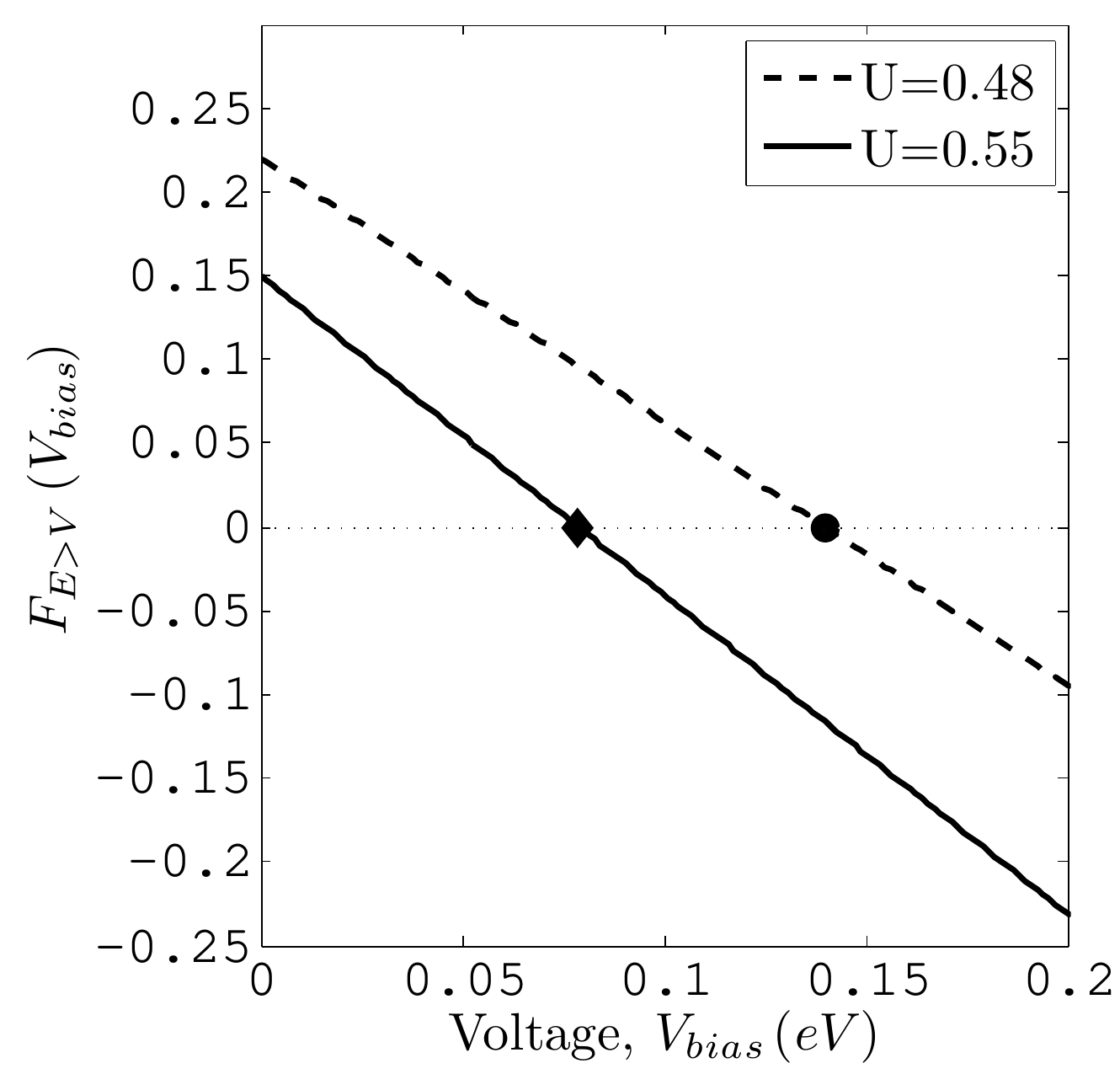} & \includegraphics[scale=0.5]{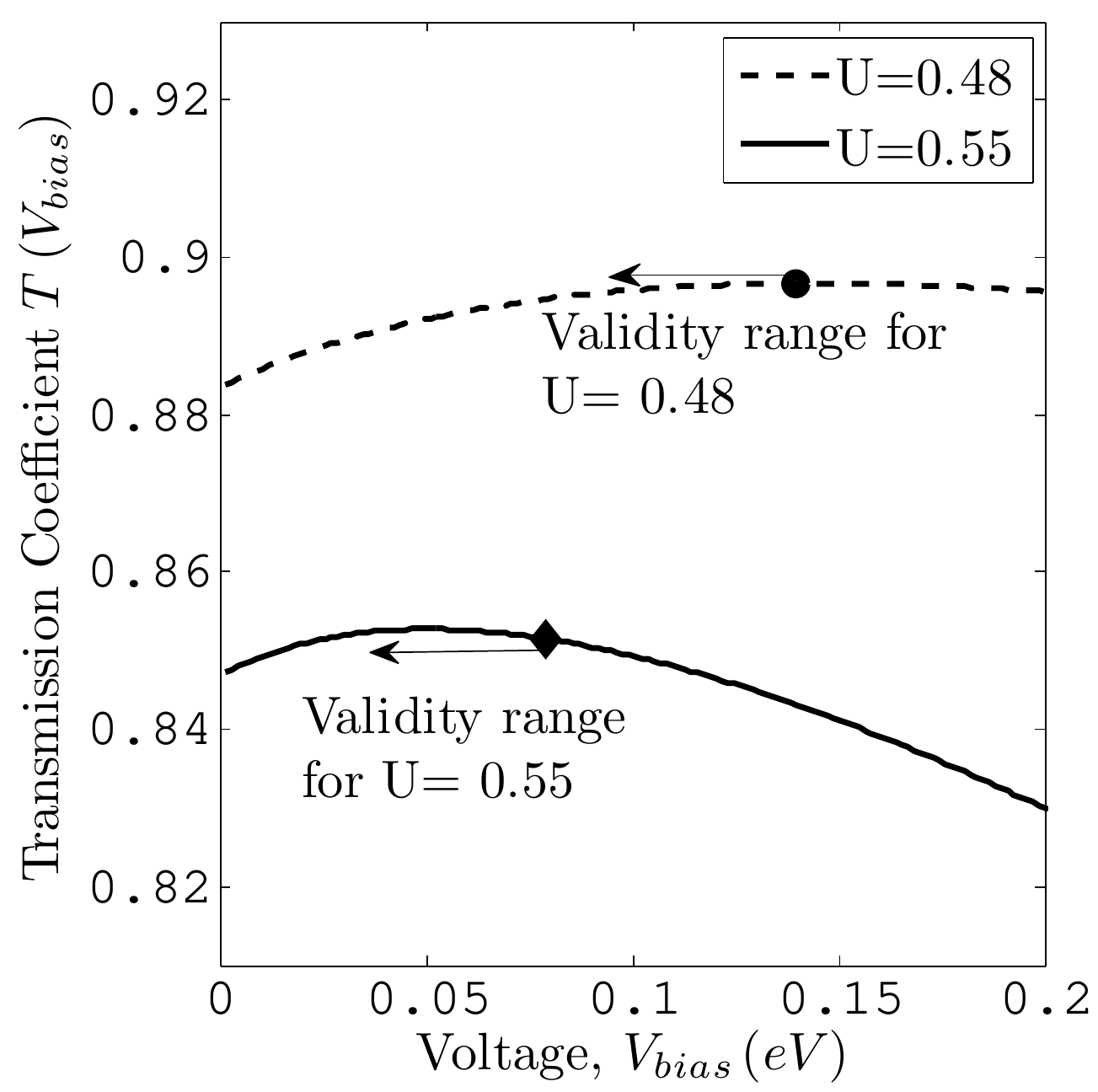}\\ 
	    \end{tabular}
\caption{{\it (Left)} Picture  of $F_{E>V}\left( V_{\text{bias}},U \right) $ for  $U=0.48$eV (dashdotted line) and $U=0.55$eV (continuous line). {\it (Right)} Transmission coefficient  $T\left( V_{\text{bias}},U \right) $ for $U=0.48$eV (dashdotted line) and $U=0.55$eV (continuous line). Range of validity for WKB method is highlighted. In both pictures, effective mass is $m=0.07 m_0$, where $m_0$ is the bare electron mass, electron's energy is $E=0.7$eV and the thickness barrier is $1$nm.}
\label{validity2}
\end{figure}

The condition for the validity of WKB method and therefore for formulas \eqref{coef_trans_WKB_E_mayor} and \eqref{coef_trans_WKB_E_menor} is that the change in the potential energy over the decay length be smaller than the magnitude of the kinetic energy (see, for instance, \cite[p. 483]{Wang_book}). For the case $E>V\left( x\right)$, this condition can be expressed as
\begin{equation}
\label{validity_WKB_E>V}
\Big\vert \hbar  \frac{dV/dx}{\kappa_2\left( x\right)}\Big\vert < \vert V\left( x\right) - E \vert\quad \forall x_1\leq x\leq x_2 \,,
\end{equation}
and for $E<V$,
\begin{equation}
\label{validity_WKB_E<V}
\Big\vert \hbar \frac{dV/dx}{\overline{\kappa}_2\left( x\right)}\Big\vert < \vert V\left( x\right) - E \vert\quad \forall x_1\leq x\leq x_2\,. 
\end{equation}
In particular, \eqref{validity_WKB_E>V} and \eqref{validity_WKB_E<V} constraint the values of $U$ and $V_{\text{bias}}$ for which the WKB methods applies. Assume that $U>V_{\text{bias}}$:  
Since $V\left( x\right)\geq U-V_{\text{bias}}$, $\kappa_2 \left( x\right) \geq \sqrt{2me\left( E-U\right)} $ for all $x_1\leq x\leq x_2$. Hence, by introducing the function
\begin{equation}
\label{F_E>V}
F_{E>V} \left( V_{\text{bias}},U\right) = \vert E-U \vert -\frac{\hbar V_{\text{bias}}}{(x_2-x_1)\sqrt{2me\vert E-U\vert} }\,,
\end{equation}
condition \eqref{validity_WKB_E>V} is satisfied whenever $F_{E>V} \left( V_{\text{bias}},U\right) >0$. Figure \ref{validity2} displays the functions $F_{E>V} \left( V_{\text{bias}},U\right)$ and its associated transmission coefficient $T\left(V_{\text{bias}},U\right)$, as given by \eqref{coef_trans_WKB_E_mayor}, for  $U=0.48$ and $U=0.55$.

\begin{figure}[h!]
\centering
	    \begin{tabular}{c c}
	    \includegraphics[scale=0.51]{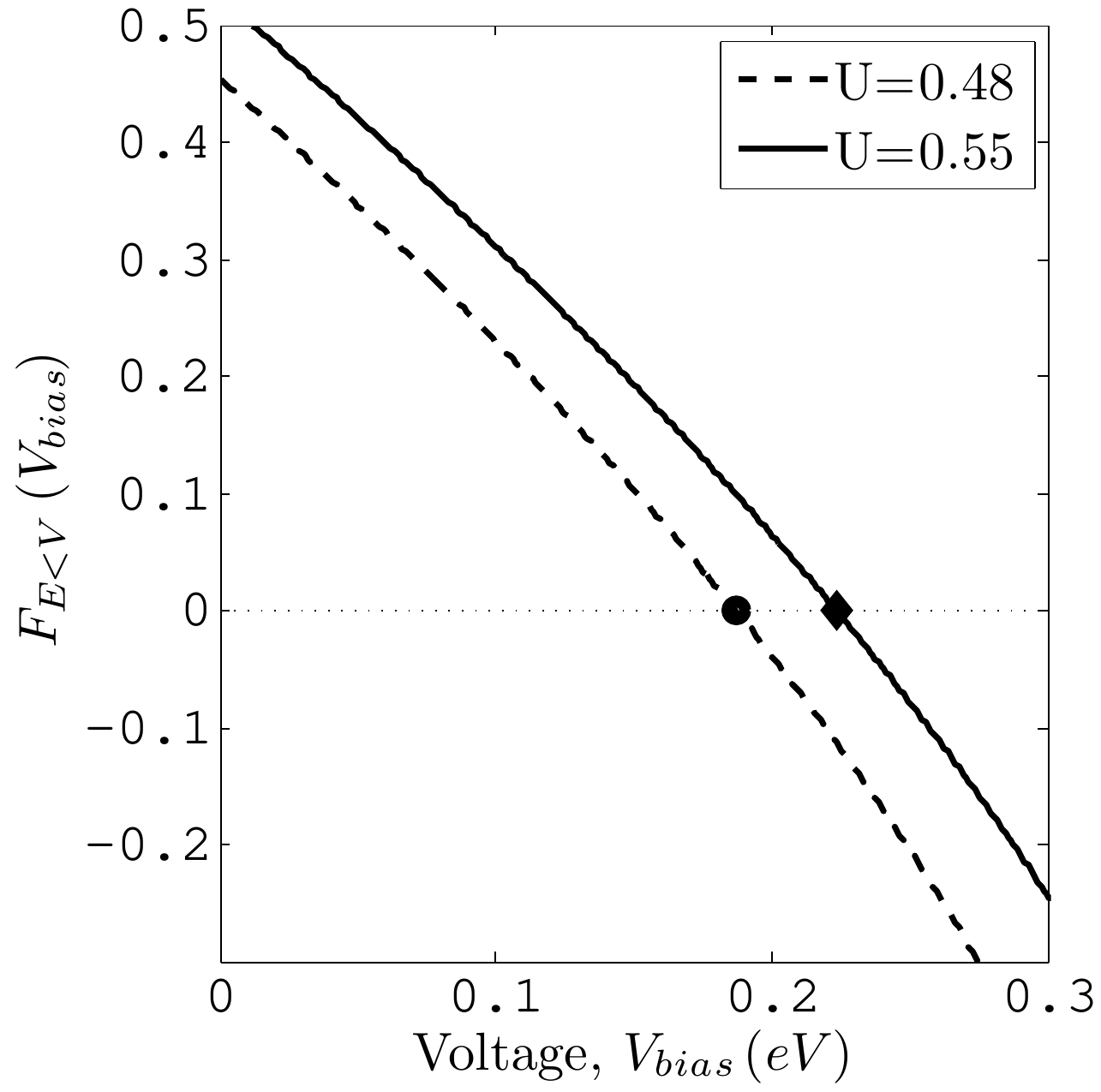} & \includegraphics[scale=0.5]{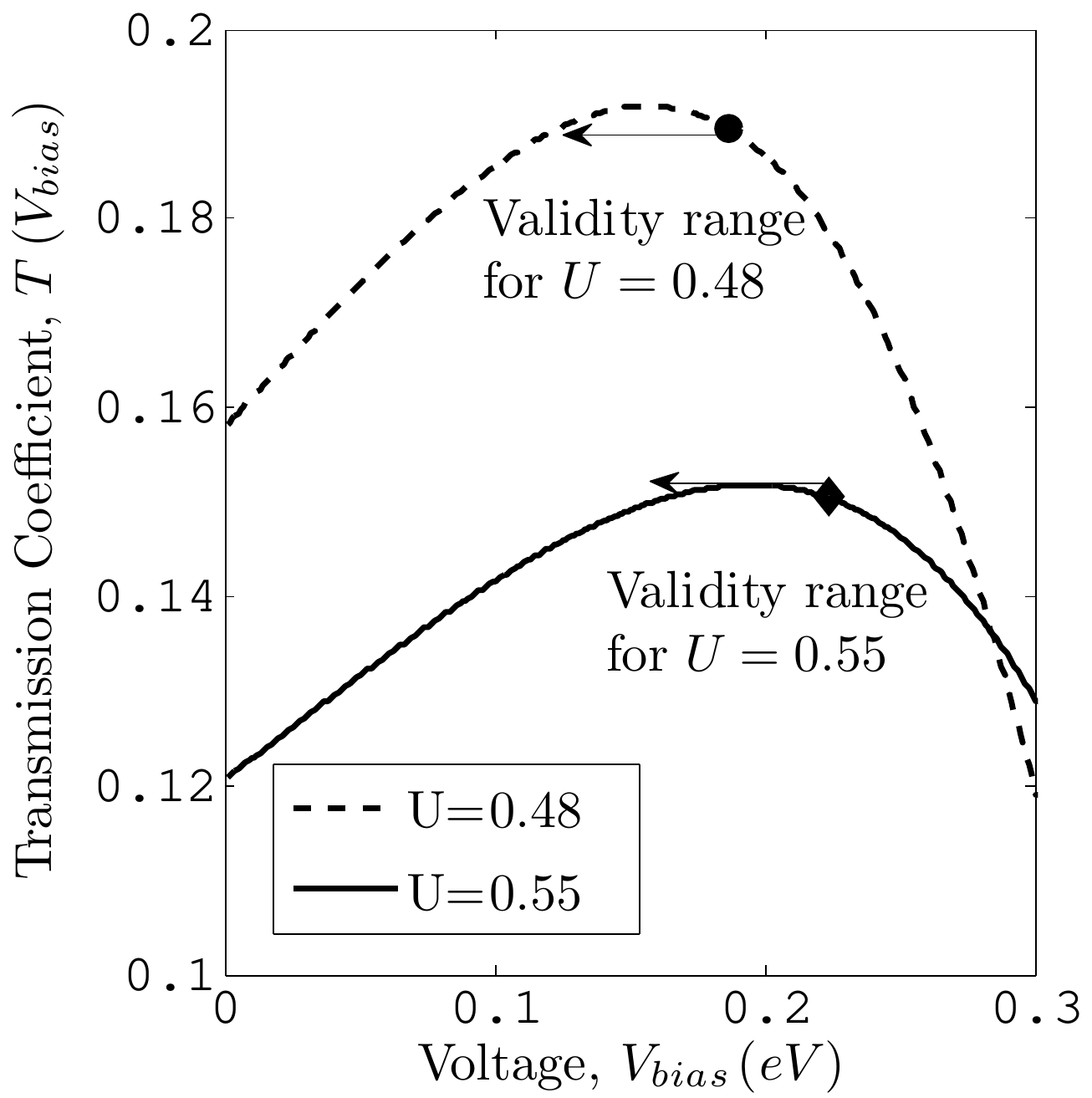}\\ 
	    \end{tabular}
\caption{{\it (Left)} Picture  of $F_{E<V}\left( V_{\text{bias}},U \right) $ for  $U=0.48$eV (dashdotted line) and $U=0.55$eV (continuous line). {\it (Right)} Transmission coefficient  $T\left( V_{\text{bias}},U \right) $ for $U=0.48$eV (dashdotted line) and $U=0.55$eV (continuous line). Range of validity for WKB method are highlighted. In both pictures, effective mass is $m=0.07 m_0$, where $m_0$ is the bare electron mass, electron's energy is $E=0.026$eV and the thickness barrier is $1$nm.}
\label{validity1}
\end{figure}

Analogously in the case $E<V$, by considering the function
\begin{equation}
\label{F_E<V}
F_{E<V} \left( V_{\text{bias}},U\right) = \vert E-U+V_{\text{bias}}\vert -\frac{\hbar V_{\text{bias}}}{(x_2-x_1)\sqrt{2me\vert E-U+V_{\text{bias}}\vert} },
\end{equation} 
condition (\ref{validity_WKB_E<V}) holds for $F_{E<V} \left( V_{\text{bias}},U\right)>0$. 
Figure \ref{validity1}  plots the functions $F_{E<V} \left( V_{\text{bias}},U\right)$ and the transmission coefficient $T\left(V_{\text{bias}},U\right)$ given in (\ref{coef_trans_WKB_E_menor} ) for two values of $U$, namely $U=0.45$ and $U=0.55$ with the energy of an electron $E=0.0.26 \text{eV}$.

\subsubsection{Case of multiple potential barriers} \label{subsub:331}

Consider now the configuration described at the beginning of Subsection \ref{Design_problem}, which is composed of $N$ potential barriers with energy given by (\ref{potential_V}). Let us denote by 
\begin{equation}\label{{V_j}}
V_j\left(x\right)=U_j-V_{\text{bias}}\frac{x-x_0}{L},\quad x_{j-1} \leq x < x_j ,
\end{equation}
the potential energy of the $j$th barrier and by
\begin{equation}\label{{V_j}}
\kappa_j\left(x\right) = \left\{
\begin{array}{ll} 
 \sqrt{2me\left( E-U_0\right)},  &  -\infty<x<x_0,\quad j=0\\
\sqrt{2me\left( E-V_j\left( x\right) \right)}, & x_{j-1}\leq x\leq x_j,\quad j=1,2,\cdots N \\
\sqrt{2me\left( E-U_{N+1}+V_{\text{bias}}\right)}, & x_N<x<+\infty, \quad j=N+1 .
\end{array}
\right.
\end{equation}
The linear relationship between the coefficients $A_0,B_0$ of the electron wave function at $x_0$ and the ones $A_N,B_N$ at $x_N$ is expressed as

\begin{equation}\label{transfer_matrices}
\begin{bmatrix}A_0 \\ B_0\end{bmatrix}=E^{-1}\left(\kappa_0,x_0\right)K^{-1}\left(k_0\right)\cdot \prod_{j=1}^N M_j\cdot K\left( \kappa_{N+1}\right) E\left(x_N,\kappa_{N+1} \right)\begin{bmatrix}A_N \\ B_N\end{bmatrix},
\end{equation}
where the form of the transfer matrices $M_j$ depend on the relationship between $E$ and $V_j$. For the cases in which WKB method may be applied, $M_j$ has been computed in the preceding section. For the spatial regions for which WKB method does not apply, appropriate connection formulae must be used. For instance, as in \cite{Levi_book,Zhang2007}, the linear potential $V_j\left( x\right)$ is approximated by piece-wise contant potential for which explicit expressions of $M_j$ are well-known \cite{Gilmore_book}. 

Finally, as in the case of a single potential barrier, the transmission coefficient $T=T\left( V_{\text{bias}},U\right)$ is obtained from (\ref{transfer_matrices}).

\subsection{Numerical Resolution Method}\label{subsec:32}

Before describing the numerical methods proposed in this paper for solving \eqref{ODP_deterministic} and \eqref{RDP}, let us briefly consider the question concerning the existence of solution for such problems.

From the explicit expressions \eqref{coef_trans_WKB_E_mayor} and \eqref{coef_trans_WKB_E_menor}, and taking into the computations in Subsection \ref{subsub:331}, it is clear that the functionals $J\left( U\right)$ and $J_{\alpha}\left( U\right)$, which have been considered in problems \eqref{ODP_deterministic} and \eqref{RDP}, respectively, are smooth (of regularity class $C^{\infty}$). In addition, the set of admissible designs $\mathcal{U}_{\text{ad}}=\left\{ U=\left( U_1,\cdots,U_N\right)\in\mathbb{R}^N : U_L\leq U_j\leq U_L,\quad 1\leq j\leq N\right\}$ is a compact set of $\mathbb{R}^N$. As a consequence, the following existence result holds.
\begin{theorem}
\label{existence}
Problems \eqref{ODP_deterministic} and \eqref{RDP} have, at least, one solution.
\end{theorem}

The nonlinear mathematical programming problem \eqref{ODP_deterministic} is  standard  and may be solved by several methods, typically by a gradient-based  method. In this paper, a subspace trust region method, which is based on the interior-reflective Newton method, as proposed  in \cite{Col94,Col96}, is used. This algorithm is implemented in the MatLab constrained optimization routine {\it fmincon}.  

More challenging is the robust optimal design problem \eqref{RDP}. The brute--force sampling Monte Carlo is the most commonly  method used to solve this kind of problems. However, for smooth (with respect to the random parameter) functions, sparse grid, stochastic collocation methods are able to keep the same accuracy as Monte Carlo and, in addition, are computationally much more efficient \cite{nobile}. In our case, again from \eqref{coef_trans_WKB_E_mayor} and \eqref{coef_trans_WKB_E_menor} it is not hard to show that $J\left( U,\omega\right)$, as defined in \eqref{random_cost}, is analytic with respect to the random variable $X\left( \omega\right)$. For this reason, we propose an adaptive, isotropic, sparse grid, stochastic collocation method to approximate both, the cost functional $J_{\alpha}\left( U\right)$ and its gradient. Then, the robust  optimal design problem \eqref{RDP} can be solved as in the deterministic case. 

In order to explain the method for approximating integrals in the random domain used in this paper, let us first introduce some notation. By $\left( \Omega, \mathcal{F},\mathbb{P} \right)$ we denote a complete probability space. $\Omega$ is the set of outcomes, $\mathcal{F}$ is the $\sigma$-algebra of events and $\mathbb{P}:\mathcal{F}\rightarrow\left[ 0,1\right]$ is a probability measure. $\Gamma_j = X_j\left( \Omega\right) $, $1\leq j\leq N$, are the images spaces of the sample space $\Omega$ through the real-valued random variables $X_j$ considered in  \eqref{uncertain_manufacturing}, and $\Gamma = \prod_{j=1}^N \Gamma_j$ is the product space. Assuming that the distribution measure of $X\left( \Omega\right)$ is absolutely continuous with respect to the Lebesgue measure, there exists a joint probability density function $\rho : \Gamma\rightarrow\mathbb{R}_+$ for $X = \left( X_1,\cdots, X_N\right)$. Hence, $\left( \Omega, \mathcal{F},\mathbb{P} \right)$ is mapped to $\left( \Gamma, \mathcal{B}, \rho\left( z\right)\, dz\right)$, where $\mathcal{B}$ is the $\sigma$-algebra of Borel sets on $\Gamma$, and $dz$ is the Lebesgue measure. Finally, the expectation and variance of $J\left( U,\omega \right)$ take the form
\begin{equation}
\label{mean_J}
\mathbb{E}\left[ J\left( U,\cdot\right)\right] = \int_{\Gamma}J\left( U, z \right) \rho (z)\,dz,
\end{equation}
\begin{equation}
\label{var_J}
\text{Var}\left[ J\left( U,\cdot\right)\right] = \int_{\Gamma}J^2\left( U, z \right) \rho (z)\,dz - \left( \int_{\Gamma}J\left( U, z \right) \rho (z)\,dz\right)^2 .
\end{equation}
Following \cite{nobile,smolyak}, the isotropic sparse grid of sampling quadrature nodes is defined as follows. Starting from an integer $\ell$ (called the level), the index set 
\[
\mathbb{I}\left( \ell,N\right) = \left\{ i=\left( i_1,\cdots,i_N\right)\in\mathbb{N}_{+}^N : \sum_{n=1}^N \left( i_n-1\right)\leq \ell \right\}
\]
is considered, with $\mathbb{N}_+ = \left\{ 1,2,3,\cdots \right\}$. The level $\ell$ determines the number of collocation points $R_{i_n}$ in the $n$th stochastic direction, which for the case of Smolyak rule is given by
\[
R_{i_n} =\left\{\begin{array}{ll} 
1,  & \text{ for } i_n=1 \\
2^{i_n-1}+1, & \text{ for } i_n>1.
\end{array}
\right.
\]
Smolyak quadrature rule applied to a generic function $ F: \Gamma\rightarrow \mathbb{R}$ gives
\begin{equation}\label{smolyakrule}
\int_{\Gamma} F\left( z\right) \rho\left( z\right)\, dz \approx  \displaystyle\sum_{{ i}\in {\mathbb{I} }\left( \ell,N\right) } \left(\Delta^{i_1}\otimes \dots \otimes \Delta^{i_N}\right) F
=  \displaystyle\sum_{r_1=1}^{R_{i_1}}\cdots\sum_{r_N=1}^{R_{i_N}}F\left( z_1^{r_1},\cdots,z_N^{r_N} \right)w_1^{r_1}\cdots w_N^{r_N},
\end{equation}
where $\Delta^{i_n}=\mathcal{Q}^{i_n}-\mathcal{Q}^{i_n-1}$, with $\mathcal{Q}^0=0$, is a quadrature rule in which the coordinates $z_n^{r_n}$ of the nodes are the same as those for the 1D quadrature formula $\mathcal{Q}^{i_n}$ and its associated weights $w_n^{r_n}$ are the difference between those for the $i_n$ and $i_n-1$ levels. 

It remains to analyze the question on how to properly choose the quadrature level $\ell$. For this:
\begin{enumerate}[(1)]
\item A positive, large enough, integer $\overline{\ell}$, and a tolerance level $0<\varepsilon\ll 1$ are fixed. 
\item The first and second order statistical moments of $J\left( U,z\right)$, as given by \eqref{mean_J} and the first term in the right-hand side of \eqref{var_J},  are approximated by using \eqref{smolyakrule}, with level $\overline{\ell}$. These two approximations, which are denoted by $\mathcal{M}_{1,\overline{\ell}} \left( J\left( U \right)\right)$ and  $\mathcal{M}_{2,\overline{\ell}} \left( J\left( U \right)\right)$, respectively, play the role of {\it enriched or reference} values for the exact values of the first two statistical moments of $J\left( U,z\right)$, which, obviously, cannot be explicitly computed. 
\item Finally, the level $\ell$ is linearly increased from $\ell =1$ to $\ell_{\text{opt}}<\overline{\ell}$ until the stopping criterion
\begin{equation}
\label{compu_error_final}
\max\left\{\frac{\vert \mathcal{M}_{1,\overline{\ell}} \left( J\left( U \right)\right)-\mathcal{M}_{1,\ell} \left( J\left( U \right)\right)\vert}{ \mathcal{M}_{1,\overline{\ell}} \left( J\left( U \right)\right)} \quad , \quad \frac{\vert \mathcal{M}_{2,\overline{\ell}} \left( J\left( U \right)\right)-\mathcal{M}_{2,\ell} \left( J\left( U \right)\right)\vert}{ \mathcal{M}_{2,\overline{\ell}} \left( J\left( U \right)\right)} \right\} \leq \varepsilon,
\end{equation}
is satisfied. Here $\mathcal{M}_{1,\ell} \left( J\left( U \right)\right)$ and $\mathcal{M}_{2,\ell} \left( J\left( U \right)\right)$ denote, respectively, approximations, by using \eqref{smolyakrule} with level $\ell$, of the first two statistical moments of $J\left( U,z\right)$. If \eqref{compu_error_final} is not satisfied for the tolerance $\varepsilon$ and the initial $\overline{\ell}$, then the reference level $\overline{\ell}$ is increased. 
\end{enumerate}
For more details on this adaptive algorithm, including its convergence, we refer the reader to \cite{nobile} and the references therein.

\section{Numerical Simulations}

In this section, numerical results for problems \eqref{ODP_deterministic} and \eqref{RDP} are presented and discussed. In all  experiments, a four layers device is considered of the same thickness ($1$nm), so that the total length is $L=4$nm. The desired linear transmission coefficient is
\begin{equation}
T_0\left( V_{\text{bias}}\right) = 0.00002  V_{\text{bias}} + 0.0000099 , \quad 0\leq V_{\text{bias}}\leq 0.25.
\end{equation}
Quadratic and square root transmission coefficients may be treated analogously.  The design is based on $10$ equally  spaced bias voltages. Hence, $V_i=i\frac{0.25}{10}$, $1\leq i\leq 10$. The electron mass is $m=0.07\times m_0$, with $m_0=9.10939 \times 10^{-31}$ Kg (which is appropriate for an electron in the conduction band of $\text{Al}_{\xi}\text{Ga}_{1-\xi}\text{As}$), its energy is $E = 0.026$ eV, and its charge $e=1.602\times 10^{-19}$ C. Planck's constant is  $\hslash=1.05457\times 10^{-34}\, J\cdot s$. The lower and upper bounds for the design variable are taken as $U_L=0.7$ eV and $U_H=1.7$ eV, respectively. 

The goal of this section is twofold: On the one hand, it is aimed at analyzing the differences that may occur when using exact gradients or numerical gradients in the optimization algorithm. On the other hand, we want to analyze the influence of manufacturing uncertainties on the computed designs.
We deal with these issues in the following subsections.

\subsection{Exact versus numerical gradient} \label{subsec:4.1}

In this experiment, the deterministic problem  \eqref{ODP_deterministic} is solved, by using the 
MatLab routine {\it fmincon}, in the cases where: (a) The exact gradient is provided, as computed
from the explicit expression for the cost functional $J\left( U\right)$ in Section \ref{sec:2}, and 
(b) the gradient is numerically computed by using finite differences.  In both cases, the algorithm 
is initiated with $U_j^0=0.7$ eV, $1\leq j\leq 4$, and the stopping criterion, provided by the MatLab routine {\it fmincon}, is fixed to $10^{-15}$.

First column in Table \ref{table_1} displays results for the values of the cost functional after convergence of the algorithm. The optimal energy potentials are showed in the remaining columns. Inspection of Table \ref{table_1} reveals (as expected) that performance increases by using the exact gradient. Precisely, the value of the objective function obtained by using the optimal design as computed with the exact gradient improves in about $69.4\%$ the corresponding one obtained via the numerical gradient. 

\begin{table}[h!]
\caption{Results for \eqref{ODP_deterministic} using exact {\it (first row)} and numerical gradient {\it (second row)}. }
\label{table_1} 
\vspace{-0.3cm}  
\begin{center} 
\begin{tabular}{lccccc}
\hline\noalign{\smallskip}
 Gradient &  $J\left( U\right)$ &   $U_1$ & $U_2$ & $U_3$ & $U_4$   \\
\noalign{\smallskip}\hline\noalign{\smallskip}
Exact & $1.43 \times 10^{-12} $ & $0.70$ & $1.31$ & $1.54$ & $0.70$  \\
Numerical & $4.68 \times 10^{-12}$ & $1.03$ & $0.96$ & $0.96$ & $1.02$   \\
\noalign{\smallskip}\hline
\end{tabular}
\end{center}
\end{table}

\subsection{Deterministic design versus  design under uncertainty} \label{subsec:4.2}

Manufacturing uncertainties are modelled by random variables uniformly distributed in $\left[ -a, a\right]$, i.e., $X_j=\mathcal{U}\left(-a,a\right)$ for all $j=1,2,3,4$. As an illustration, the cases $a=0.05$ and $a=0.2$ are considered. As in the preceding example, the algorithm is initiated with $U_j^0=0.7$ eV, $1\leq j\leq 4$, and the stopping criterion is fixed to $10^{-15}$.  

\begin{figure}[h!]
\includegraphics[scale=0.5]{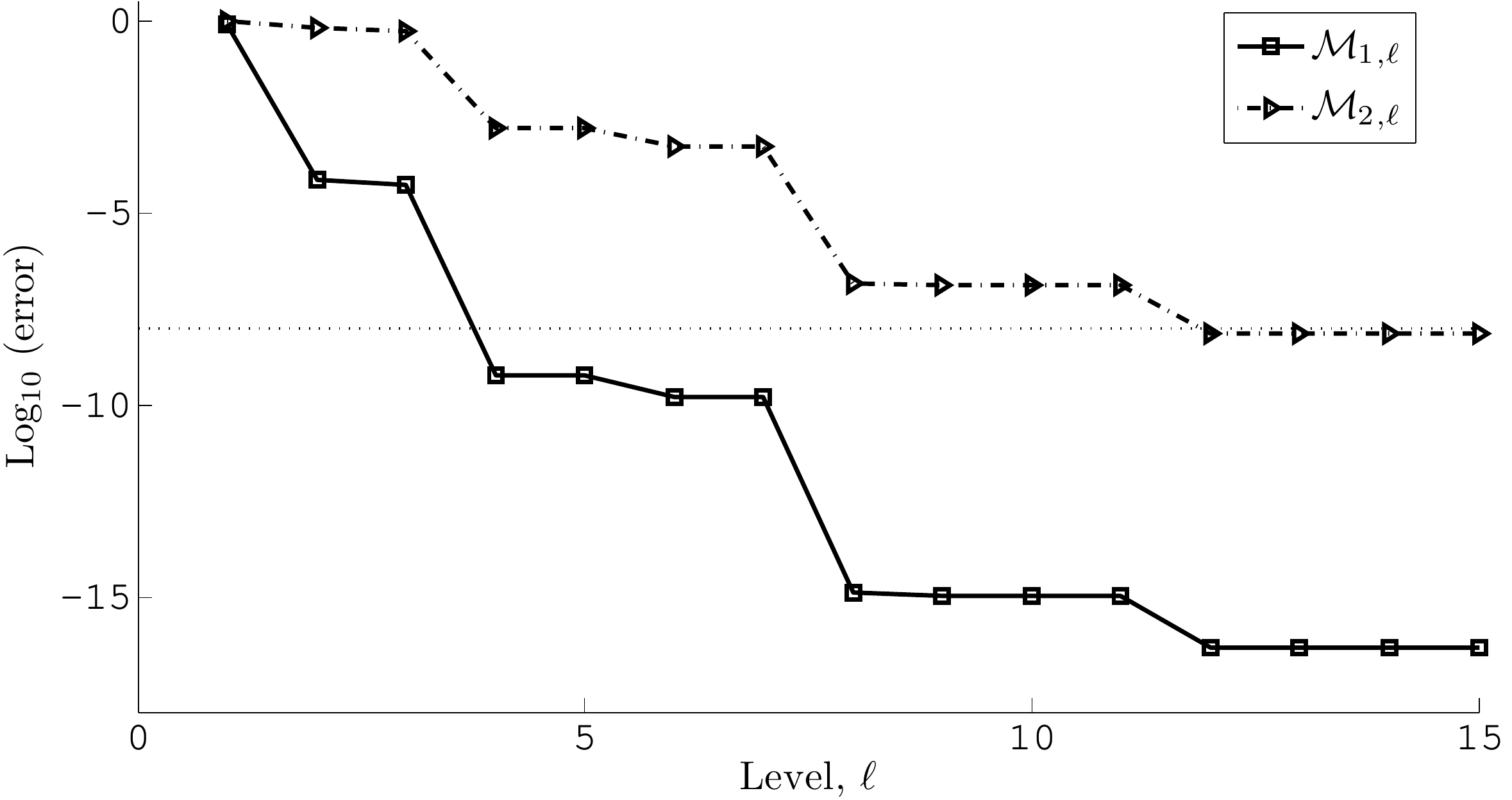}
\centering
\caption{Case $a=0.05$. Rates of convergence of the isotropic sparse grid algorithm. In the vertical 
axis, $\log_{10}\left( \text{error}\right)$ represents the two terms for relative error considered in 
\eqref{compu_error_final}. Precisely, the continuous line corresponds to relative error for the first 
order moment of $J\left( U,\cdot\right)$, and the dashed--dotted line is its second moment. The 
horizontal dashed line represents the prescribed accuracy level in the stopping criterion 
\eqref{compu_error_final}.}   
\label{fig:rate_convergence}
\end{figure}

\noindent\textbf{Case 1: $a=0.05$},  which represents $5\%$ of error in manufacturing each one of the potentials $U_j$.\\
The algorithm described in Subsection \ref{subsec:32} has been implemented by using the  Sparse Grids Matlab kit 15.8 (see \url{http://csqi.epfl.ch} and \cite{nobile}). The stopping criterion (\ref{compu_error_final}), with $\varepsilon = 10^{-7}$, is satisfied for $\overline{\ell}=20$ and $\ell_{\text{opt}}=15$, which corresponds to $895$ collocation nodes in the random domain $\Gamma = \left[ -0.05, 0.05\right]^4.$ Figure \ref{fig:rate_convergence} displays the rates of convergence of the isotropic sparse grid algorithm. 

Table \ref{table_2}  shows the results, after convergence of the algorithm, for the expectation and 
the standard deviation of the cost functional $J\left( U,\omega\right)$ given by \eqref{random_cost}. 
As expected, the optimal design in mean ($\alpha =0$) provides a solution which  reduces the impact of 
manufacturing errors in comparison with the deterministic approach (see first and second rows in the 
first column of Table \ref{table_2}). It is also observed that increasing the value of the weighing 
parameter $\alpha$ reduces the dispersion of the computed designs. These results are more significant 
when the level of uncertainty increases, as pointed out in Table \ref{table_3}. 

\begin{table}[h!]
\caption{Case $a=0.05$. Mean {\it (first column)} and standard deviation {\it (second column)} of the  functional (\ref{random_cost}) for the optimal deterministic design {\it (first row)},  the optimal design in mean {\it (second row)}, and the optimal design for $\alpha =10^{12}$ {\it (third row)}.  }
\label{table_2}
\vspace{-0.3cm}  
\begin{center} 
\begin{tabular}{lcccccc}
\hline\noalign{\smallskip}
 Design &  $\mathbb{E}\left( J\left( U,\cdot\right)\right)$ &   $\text{std }\left( J\left( U,\cdot\right)\right)$  &   $U_1$ & $U_2$ & $U_3$ & $U_4$   \\
\noalign{\smallskip}\hline\noalign{\smallskip}
Deterministic & $1.737\times 10^{-11}$ & $2.143\times 10^{-11}$ & $0.70$ & $1.31$ & $1.54$ & $0.70$  \\
$\alpha =0$ & $1.664\times 10^{-11}$ & $1.920\times 10^{-11}$ & $0.95$  & $0.90$ & $1.60$  & $0.70$ \\
$\alpha = 10^{12}$ & $1.908\times 10^{-11}$ & $1.845\times 10^{-11}$ & $1.14$ & $0.83$ & $1.05$  & $0.97$ \\
\noalign{\smallskip}\hline
\end{tabular}
\end{center}
\end{table}

\noindent\textbf{Case 2: $a=0.2$},  which corresponds to $20\%$ of noise. The same procedure as in the preceding case has been applied. in this case, (\ref{compu_error_final}) is satisfied, with $\varepsilon = 10^{-2}$, for $\overline{\ell}=20$ and $\ell = 16$, which corresponds to $1212$ collocation nodes.

\begin{table}[h!]
\caption{Case $a=0.2$. Mean {\it (first column)} and standard deviation {\it (second column)} of the   cost functional (\ref{random_cost}) for the optimal deterministic design {\it (first row)},  the optimal design in mean {\it (second row)}, and the optimal design for $\alpha =10^{10} $ {\it (third row)}. Columns $3$ to $6$ show the corresponding optimal potentials. }
\label{table_3} 
\vspace{-0.3cm}  
\begin{center} 
\begin{tabular}{lcccccc}
\hline\noalign{\smallskip}
 Design &  $\mathbb{E}\left( J\left( U,\cdot\right)\right)$ &   $\text{std }\left( J\left( U,\cdot\right)\right)$  &   $U_1$ & $U_2$ & $U_3$ & $U_4$   \\
\noalign{\smallskip}\hline\noalign{\smallskip}
Deterministic & $3.48\times 10^{-10}$  & $7.17\times 10^{-10}$ & $0.70$ & $1.31$ & $1.54$ & $0.70$  \\
$\alpha =0$ & $2.12 \times 10^{-10}$ & $2.66 \times 10^{-10}$
 & $1.19$  & $0.82$ & $1.21$ & $0.92$ \\
$\alpha = 10^{10}$ & $2.35\times 10^{-10}$& $2.15\times 10^{-10}$ & $1.20$ & $0.88$ &  $1.07$ & $1.05$\\
\noalign{\smallskip}\hline
\end{tabular}
\end{center}
\end{table}

The same qualitative results as in the preceding case are observed in Table \ref{table_3}. However, since in this case the level of manufacturing uncertainties is higher, the differences of corresponding solutions are much more significant than in the preceding case. Indeed, for $20\%$ of noise, the reduction in the mean value of $J\left( U,\cdot\right)$ for the mean optimal value ($\alpha =0$), in comparison with the deterministic optimal design, is of the order of $39\%$. For the case of $5\%$ of noise, this reduction is of the order of $4.2\%$. The decreasing if the standard deviation of the computed designs is also much more significant in the current case than in the case of $5\%$ of manufacturing noise.   

\section{Conclusions}
We have addressed the problem of determining optimal designs of nanoelectronic devices whose physical behavior is governed by the time--independent Schr\"odinger equation. Two situations are considered: A deterministic version of the problem, and the more realistic case where manufacturing uncertainties are accounted for. In both cases, the corresponding optimal design problems are formulated as the minimization of a least--squares performance metric. In the stochastic case, the variance of the random metric is incorporated in the cost functional as a measure of robustness. An  explicit expression for that metric is computed using a semi--classical approximation. 
Having at our disposal an analytic expression for the cost functional has the 
following advantages:
\begin{enumerate}[(a)]
\item At the theoretical level, it is easily deduced that the cost functional is smooth (of regularity 
class $C^{\infty}$) with respect to the design variables. As a consequence, the existence of solutions 
for both optimization problems is proved.
\item In the deterministic case, an explicit expression for the gradient of the cost functional is obtained. Numerical simulations in Subsection \ref{subsec:4.1} show the relevance of this issue, at least in the specific problem considered in this work, when gradient--based minimization algorithms are used as the numerical resolution method. 
\item In the stochastic optimization problem, it is also deduced that integrands, which appear in the considered cost functional, are analytic with respect to the random vector parameter. Thus,
stochastic collocation methods (which possess an exponential rate convergence with respect to the number of sampling points and are, computationally, much more efficient than the classical brute-force Monte-Carlo method) are preferred.      
\end{enumerate}

The results obtained show an improvement of the accuracy in the linear response characteristic
of about $69{.}4\%$ over previous, brute--force, approaches. Moreover,
the robustness of the design is manifest even under weight values of $\alpha =10^{12}$ in the variance
(physically, as seen in in Table \ref{table_2}, this would amount to pass from a design using Germanium, with a band gap of
$\simeq$ 0.7eV, to one using Silicon, whose band gap is $\simeq$ 1.1eV, so that extreme case is still physically feasible).   

As noticed along the paper, the semi--classical approach used in this work has the drawback that constraints the values of the applied bias voltage and those of the local energy potentials. Accordingly, the whole range of possible energies and potential profiles may be covered by using the approach proposed in this work in combination with appropriate connection formulas, for instance, the ones presented in \cite{Levi2010}. 

\vspace{0.3cm}

\textbf{Acknowledgements.} OM was supported by grant number $726714$ from CONACyT (Consejo Nacional de Ciencia y Tecnolog\'{\i}a, Mexico), under program Movilidad en el extranjero (291062). FP was supported by  projects  DPI2016-77538-R from Ministerio de Econom\'{\i}a y Competitividad (Spain)
and 19274/PI/14 from Fundaci\'{o}n S\'{e}neca (Agencia de Ciencia y Tecnolog\'{i}a de la Regi\'{o}n de Murcia (Spain)).  JAV was supported by a CONACyT project CB-179115.


\end{document}